\documentstyle[preprint,aps]{revtex}
\def\beq{\begin{equation}}
\def\eeq{\end{equation}}
\def\winf{\ W_{1+\infty}\ }
\begin{document}
\preprint{MPI-Ph/93-75 \quad DFTT 65/93}
\title
{Classification of Quantum Hall Universality Classes
by $\ W_{1+\infty}\ $ symmetry}

\author{Andrea Cappelli}
\address{I.N.F.N. and Dipartimento di Fisica,
         Largo E. Fermi 2,
         I-50125 Firenze, Italy}
\author{Carlo~A.~Trugenberger}
\address{Max-Planck-Institut f\"ur Physik,
         F\"ohringer Ring 6, D-80805 Munich, Germany}
\author{Guillermo~R.~Zemba}
\address{I.N.F.N. and Dipartimento di Fisica Teorica,
          Via P. Giuria 1,
         I-10125 Torino, Italy}

\maketitle
\begin{abstract}
We show how two-dimensional incompressible quantum fluids and their
excitations can be viewed as $\winf$ edge conformal field theories,
thereby providing an algebraic characterization of incompressibility.
The Kac-Radul representation theory of the $\winf$ algebra leads then
to a purely algebraic complete classification of hierarchical quantum Hall
states, which encompasses all measured fractions.
Spin-polarized electrons in single-layer devices can only have Abelian
anyon excitations.
\end{abstract}


\narrowtext

The central idea in Laughlin's theory~\cite{Lau} of the
quantum Hall effect \cite{Pra}
is the existence of {\it two-dimensional incompressible
quantum fluids} at specific rational values $\rho =\nu eB/hc $ of the
electron density ($B$ being the external magnetic field).
These are macroscopical quantum states with uniform density
which are rendered particularly stable by an energy gap.
We can think of them semiclassically as {\it droplets of liquid}
without density waves.
The absence of density waves causes the longitudinal conductivity to vanish
while the Hall conduction is realized as an overall rigid translation of the
droplet, giving the Hall conductivity ${\sigma}_{xy} = \nu e^2/h\ $ .

Our program has been to characterize this picture by a symmetry principle
and derive accordingly the {\it effective field theory} for low
energy excitations.
Indeed, the {\it universality} observed in experiments calls for an effective
theory approach at long distances, while the extreme precision of the
rational values of $\nu$ suggests that dynamics is constrained
by {\it symmetry}.
Both facts suggest an analogy with two-dimensional critical phenomena, which
are classified by conformal field theories \cite{Gin}.

In earlier works~\cite{Cap1},\cite{Cap2}, we already identified
the algebra $\winf$ of {\it quantum area-preserving diffeomorphisms}
\cite{Shen}, as the relevant symmetry
characterizing the simplest incompressible quantum fluids.
Here we construct a complete, purely algebraic classification of
all universality classes of incompressible quantum fluids as $W_{1+\infty}$
conformal field theories on the edge of the droplet.
In addition to obtaining an algebraic characterization of incompressibility,
the recently constructed \cite{Kac} representation theory of $\winf$
allows for the identification of the quantum numbers of the excitations.

The idea of our construction is simple \cite{Cap1}.
We start by recalling that all possible configuration of {\it classical}
two-dimensional incompressible fluids are connected by
{\it area-preserving diffeomorphisms}
of the two-dimensional plane $\ (z=x+iy,\ {\bar z}=x-iy)\ $.
Infinitesimal transformations of the distribution function
$\rho (z,{\bar z})$ of the fluid are generated by
the complete basis
${\cal{L}}^{(cl)}_{n,m}= {\rho_o}^{(n+m)/2} z^{n} {\bar z}^{m}$
via the Poisson brackets
$\{ f ,g \}_{PB}\equiv {i/\rho_o}
\left( \partial f {\bar\partial} g - {\bar\partial} f \partial g \right)$,
with $\partial \equiv {\partial}/{\partial z}$,
${\bar\partial} \equiv {\partial}/{\partial{\bar z}}$ and
$\rho_o$ the average density.
The generators ${\cal{L}}^{(cl)}_{n,m}$ satisfy the classical
algebra
\begin{equation}
\{ {{\cal{L}}^{(cl)}_{n,m}},{{\cal{L}}^{(cl)}_{k,l}}
\}_{PB}\ = -i \left( mk-nl \right )
{{\cal{L}}^{(cl)}_{n+k-1,m+l-1}}\ ,
\label{area}
\end{equation}
which is usually denoted \cite{Shen} by $w_{\infty}$.
All classical ``small excitations'' around a fluid ground state
describing a droplet with uniform density $\rho_o$ are given by the basis
$\delta\rho =c_{n,m} \{{\cal{L}}^{(cl)}_{n,m}, \rho\}_{PB} $.
As can be checked explicitly~\cite{Cap1}, they correspond to density waves
localized on the sharp boundary of the classical droplet.

Given this fundamental role of $w_{\infty}$ in the description of
classical incompressible fluids, it is to be expected that its
{\it quantum extension} $W_{1+\infty}$, obtained by the substitution
$i\hbar\{\ ,\ \}_{PB} \to [\ ,\ ]$ $(z\to z , {\bar z}\to\partial)$,
plays a corresponding role in the
physics of {\it quantum incompressible fluids}.

That this is indeed so can be checked explicitly for the simplest
example of a quantum incompressible fluid, the fully filled Landau
level \cite{Cap1}\cite{Saki}.
As was pointed out by Stone~\cite{Stone}, the incompressibility
of this state follows solely from the fact that it behaves as a
filled Fermi sea in the coordinate plane of the electrons (as opposed
to momentum space).

The field operator for spin-polarized non-relativistic planar electrons
subject to the uniform magnetic field and confined to the first Landau level
is given in the symmetric gauge
$A_i =B/2\epsilon^{ij}x^{j}$ by
$\ \Psi (\vec{x}) = \sum _{k=0}^{\infty }\ a_k\ \psi _k(\vec{x})$,
where
\beq
\psi _k(z, \bar z) = {1\over \ell \sqrt{\pi}} {1\over \sqrt{k!}} \left(
{z\over \ell}\right)^k {\rm exp}\left(- {|z|^2\over 2\ell ^2}\right)\ ,
\label{psi}
\eeq
and we denote by $\ell =\sqrt{2/eB}$ the
magnetic length (we use units $\hbar =c=1$).
The coefficients in the above expansion are fermionic
Fock annihilators satisfying
$\left\{ a_k ,a_l^{\dag } \right \}=\delta _{kl}$ ,
with all other anticommutators vanishing.
The fully-filled Landau level is a configuration in which all angular
momentum eigenstates $\ \psi_k\ $ are occupied up to (and including) a
maximal angular momentum $L$:
$|\Omega\rangle\ =\ a^{\dag}_{0} a^{\dag}_{1} \dots a^{\dag}_{L}
|0\rangle\ $,
where $|0\rangle $ is the Fock vacuum. Since the single-particle angular
momentum eigenstates $\psi _k$ are peaked around radii $r_{k} = \ell
\sqrt{k}$, the state $|\Omega\rangle$ consists of a circular droplet of radius
approximately given by $\ell \sqrt{L}$ (the boundary is smoothed out
by quantum effects). This ground state is {\it quantum incompressible}
in the sense that compressions of the droplet are forbidden by an energy
gap. Actually, these would lower the total angular momentum
by promoting at least one electron to the next Landau level.

At the quantum level, the boundary density waves of the classical droplet
picture become ``neutral'' particle-hole excitations across the Fermi surface.
These are the edge excitations \cite{Wen}
to be included in the low-energy long-distance effective theory,
valid in the thermodynamic limit $L\to\infty$.
The idea is that to leading non-trivial order in this limit,
the filled Landau level can be substituted by an {\it infinite Dirac sea},
as we now show \cite{Cap2}.
We first pick a radius $R$
such that $\left(R/\ell\right)^2 =L+\mu, 0\le\mu <1$ and we then consider
the field operator of the first Landau level restricted to the edge $r=R$,
rewritten as follows:
\beq
\Psi\left( r=R,\theta\right)\ \equiv \left({2\over \pi\ell^2}\right)^{1/4}\
{\rm e}^{i(L+\mu)\theta}\ F_R\left(\theta\right) \ ,
\label{phirel}
\eeq
where
\begin{eqnarray}
\sqrt{R}\ F_R (\theta) \ &=& \sum_{k=-L}^{\infty}\  {c_k\over\sqrt{2\pi}}
     {\rm e}^{i(k-\mu)\theta}\ b_k \ ,\nonumber\\
c_k^2 ={\sqrt{2\pi}\over \left(L+k\right)!} &\ &
     \left({R\over\ell}\right)^{2L+2k+1}\left.{\rm e}^{-R^2/\ell^2}
      \right\vert_{R^2=(L+\mu)\ell^2}
\label{fr}
\end{eqnarray}
and we introduced the shifted operators $b_k\equiv a_{L+k}$.
The coefficients $c_k$ are approximated by
$\ c_k\sim \exp\left(-(k-\mu)^2/2L\right)\ $ for $\ L\to\infty\ $ and
provide, therefore, a smooth ultraviolet cutoff
$\ \vert k\vert <\sqrt{L}\ $ for the sum over the angular momentum $k$.
To leading order in $1/L$ we can remove this cutoff thereby obtaining
the {\it effective} relativistic chiral fermion field
\beq
F(\theta)\equiv\lim_{R\to\infty} \ \sqrt{R} \ F_R =
{1\over \sqrt{2\pi}}\ \sum_{k=-\infty}^{\infty}\
{\rm e}^{i(k-\mu)\theta}\ b_k\
\label{ftheta}
\eeq
in $(1+1)$ dimensions (Weyl fermion),
with boundary conditions determined by the ``chemical potential'' $\mu$.

Next we define the $\winf$ generators in the relativistic theory
(they can equivalently be obtained by quantization
of the $\ {\cal L}_{nm}^{(cl)}\ $ in the Landau level problem
and by the limit (\ref{ftheta}) \cite{Cap1}):
\begin{eqnarray}
V_n^j\ &=&\ \int_0^{2\pi} d\theta\
F^{\dag}(\theta)\ \ddagger {\rm e}^{-in\theta}\left(-i\partial_{\theta}
\right)^j \ddagger\ F(\theta)\ , \nonumber\\
\ &=&\ \sum_{k=-\infty}^{\infty}\ p(k,n,j;\mu)\ b^{\dag}_{k-n}\ b_k \ ,
\qquad j\ge 0\ ,
\label{vi}
\end{eqnarray}
where $\ddagger\ \ddagger $ denotes an ordering of the first-quantized
operators
$\ \exp (-i\theta)\ $ and $\ i\partial_{\theta}\ $ to be specified below,
and such that ${V^j_n}^{\dag}=V^j_{-n}$.
The coefficients $\ p(k,n,j;\mu)\ $ are $j$-th order polynomials
in $k$ whose specific form depends on the choice of ordering.
Let us now compute the algebra satisfied by the $\ V^i_n$'s.
Given the standard anticommutation rules of the fermionic operators $F$
and $F^{\dag}$, the operator part of $\ [ V^i_n ,V^j_m ]\ $ reproduces
the corresponding first quantized commutator. We thus obtain
\begin{eqnarray}
\left[ V^i_n, V^j_m \right] &=& \left(jn-im\right) V^{i+j-1}_{m+n} +
q(i,j,m,n)V^{i+j-3}_{m+n}\nonumber\\
 &+&\dots\ + \delta_{n+m,0}\ c\left(n,i,j\right)\ .
\label{walgebra}
\end{eqnarray}
The first term in the r.h.s. reproduces the classical algebra (\ref{area})
by the correspondence $\ {\cal L}_{i-n,i}\to V^i_n\ $, and
identifies (\ref{walgebra}) as the algebra $\ W_{1+\infty}\ $ of
``quantum area-preserving diffeomorphisms'' \cite{Shen}.
The second and higher operator terms in the r.h.s. arise at the quantum
level because the $V^i_n$ are polynomials in $\ \partial_{\theta}\ $.
Finally, the $c$-number terms $\ c(n,i,j)\ $ represent the relativistic
quantum anomaly. This follows from the renormalization
of the charges $\ V^i_0\ $.
Since we want to measure charges with respect to the original filled
Landau level, we adopt the standard relativistic normal ordering procedure
of writing all annihilators to the right of creators.
Thus, it can be shown \cite{Cap3} that the coefficients $\ c(n,i,j)\ $
depend only on
the first-quantized ordering $\ddagger\ \ddagger$ and on $\mu$.
In particular, they can be made diagonal, $\ c(n,i,j)=c(n)\delta^{ij}\ $,
for a specific ordering.
Independently of our explicit construction, it has been shown\cite{Radul}
in general that there is a unique central extension for the algebra
(\ref{walgebra}) in the $(1+1)$-dimensional relativistic field theory.

The simplest cases of eq.(\ref{walgebra}) are
\begin{eqnarray}
\left[ V^0_n,V^0_m\right] &=& n\ \delta_{n+m,0}\ ,\nonumber\\
\left[ V^1_n,V^0_m\right] &=& -m\ V^0_{n+m}\ ,\nonumber\\
\left[ V^1_n,V^1_m\right] &=& (n-m) V^1_{n+m} +
        {c\over 12}\left(n^3-n\right)\delta_{n+m,0}\ ,
\label{kacmoody}
\end{eqnarray}
with $c=1$ and
where $\ \mu=1/2\ $ has been chosen in order to cancel the anomaly in
the second commutator (note that eqs. (\ref{kacmoody}) are
independent of the first quantized ordering $\ddagger\ \ddagger$).
Eqs.(\ref{kacmoody}) show that $\ V^0_n\ $ and $\ V^1_n\ $ are the oscillator
and conformal modes, respectively, and the central charge is $\ c=1\ $,
as expected for a Weyl fermion.
The index $(i+1)$ is the conformal spin of the $V^i_n$ currents and
$n$ is the moding.

The quantum incompressibility of the filled Landau level in the limit $\ L\to
\infty\ $ can now be characterized by the infinite set of conditions
\beq
V^i_n\ \vert\Omega\rangle\ =\ 0,\qquad \forall\ \  n> 0\ ,i\ge 0\ .
\label{hwc}
\eeq
In mathematical terms \cite{Gin}, this equations states that the
filled Landau level is a
{\it c=1 highest-weight state} of the $\ W_{1+\infty}\ $ algebra with
weights $\ V^i_0\vert\Omega \rangle =0,\ \forall i\ge 0\ $.
Moreover, all neutral excitations generated by polynomials of $\ V^i_n\ (n<0)$
applied to $\vert\Omega\rangle$, make up a
{\it unitary irreducible highest weight representation of} $W_{1+\infty}$.

There exist also other excitations, corresponding to
quasi-holes and quasi-particles in the bulk of the droplet \cite{Lau}.
Actually, due to incompressibility any excitation corresponding to a
local density deformation in the bulk is transmitted to the edge, where
it is seen as a {\it charged} excitation.
In our algebraic language, it can be shown that these charged
excitations, together with their towers of neutral excitations,
correspond to further irreducible highest-weight representations \cite{Cap3}.
All these highest-weight representations define a
$\ W_{1+\infty}$ conformal field theory, in short, a
$\winf$ {\it theory}.
This is the theory of the free Weyl fermion in the case of the filled
Landau level discussed so far.

We are thus naturally led to characterize {\it all two-dimensional
incompressible quantum fluids with their edge excitations as
$\ W_{1+\infty}$ theories}
(actually, analogs of the conditions (\ref{hwc}) are known \cite{Cap3}
\cite{Flohr} for the
Laughlin and hierarchical fluids).
This is a powerful classification scheme, given that all highest-weight
representations of $\ W_{1+\infty}\ $ have been recently obtained by
Kac and Radul \cite{Kac}.
Specifically, we are interested in classifying two-dimensional
incompressible fluids by their {\it filling fraction} $\nu$
and by the {\it charge} and {\it spin (statistics)} quantum numbers of their
{\it quasi-particle excitations}.
These are the eigenvalues of the operators $\left(-V^0_0\right)$ and
$V^1_0 $ in the highest-weight representation
corresponding to a given excitation.
Note that the charges of bulk excitations are $\left(-V^0_0 \right)$,
with opposite sign to their edge counterparts.
Moreover, the statistics $\theta/\pi$, computed
from monodromies along the edge, is twice the spin.

The results of Kac and Radul relevant to our construction can be conveniently
rephrased in our basis as follows.
All unitary, irreducible, highest-weight representations of
$\ W_{1+\infty}$ are completely characterized by a $(m)$-dimensional
multiplet of ``charges'', the ``vector''
$\vec s$, giving the anomaly $c$ and the highest weights ${\cal Q}$
and $J$ of $\left(-V^0_0\right)$ and $V^1_0$, respectively,
\begin{eqnarray}
\vec{s}\ &\equiv &\ \{ s^I \in {\bf R}, \ I=1,\dots,m\} \ ,\quad
c=m\ ,\ m \in {\bf Z}_+\ ,\nonumber\\
{\cal Q}\ &=&\ s^1 + s^2 +\dots + s^m\ ,\nonumber\\
J\ &=&\ \frac{1}{2}\left[ (s^1)^2 + (s^2)^2 +\dots + (s^m)^2 \right]\ .
\label{cqj}
\end{eqnarray}
According to the rules of conformal field theory \cite{Gin}, a $\winf$ theory
is defined by a set of these representations which is closed under
the ``fusion algebra'', in physical terms the making of composite
excitations.
The ``fusion rule'' for $\winf$ is the addition of
charge vectors $\vec{s}$, as in the well-known case of the affine
subalgebra (\ref{kacmoody}), because
the unitary representations of these two algebras are in one-to-one
correspondence \cite{Cap3}.
Therefore, a consistent set of representations is given by a {\it lattice}
$\Gamma$ generated by $m$ ``elementary'' excitations $\ \vec{v}_i \ $:
\beq
\Gamma\ =\ \{ \vec{s}:\ \vec{s}\ =\ \sum_{i=1}^{m} n_i \vec{v_i},\
n_i \in {\bf Z},\ i=1,\dots,m \}\ .
\label{lat}
\eeq

These results tell us immediately that a level $c=m$
incompressible quantum fluid is a composite with $m$ edges. Indeed, a
$c=m$ $\ W_{1+\infty}\ $ representation looks like a
superposition of $m$ $c=1$ $\ W_{1+\infty}\ $ representations, for
which the charges can be written as
$V^0_n = \sum_{I=1}^{m}\left( V^0_n \right)^I $.
The crucial point is that the $\left( V^0_n\right)^I$
need not be the appropriate basis describing the {\it physical charge
operators}
$\rho^I_0$ on the $m$ edges. These are generically given by
\begin{equation}
\rho^I_n\ =\ \sum_{J=1}^{m}\ \Lambda_{IJ}\ \left( V^0_n \right)^{J}\ ,
\qquad \Lambda \in {\rm GL}(m,{\bf R})\ .
\label{physch}
\end{equation}
The matrix $\Lambda$ is operationally determined by the chiral couplings of
the physical charge densities to independent electromagnetic fields $A^I$
on the edges,
\begin{eqnarray}
H=H_0 + H_{e.m.}&=& \sum_{I=1}^m \left(
{v^I\over R^I}\left(V_0^1\right)^I +
\int_{0}^{2\pi} d\theta \rho^{I} A^{I} \right) ,\nonumber\\
\rho^{I} \left( \theta - {\frac{v^I}{R^I}}t \right)\ &\equiv &\
{\frac{1}{2\pi}} \sum_{n=-\infty}^{\infty} \rho^{I}_{n}
\ {\rm e}^{i n (\theta - tv^I/R^I)}\ ,\nonumber\\
A^{I} &\equiv & \left( A_0 + v^I  A_\theta \right)\vert_{r=R^I}\ .
\label{coup}
\end{eqnarray}
In eq.(\ref{coup}), $H_0$ describes the relativistic chiral dynamics
of the edge modes provided by the confining boundary potential \cite{Wen},
which  is parametrized phenomenologically by the effective
velocities $v^I$ on the $m$ edges at radii $R^I$.
Here we consider only chiral incompressible fluids, for which all
$v^I$ have the same sign; the generalization to both chiral and antichiral
edges is straightforward.

The {\it chiral anomaly} generated by the coupling (\ref{coup})
is completely fixed by the current algebra
(\ref{kacmoody}) and eq.(\ref{physch}).
It determines the charges created by the electric fields $E^I$
out of the vacuum:
\begin{eqnarray}
\left.\Delta \left( V^0_0 \right)^{I} \right\vert_{t=-\infty}^{t=\infty}
&\equiv & -\sum_{J} \Lambda_{JI}
\int_{-\infty}^{+\infty} dt \int_{0}^{2\pi}
{\frac{d\theta}{2\pi}} E^J  \nonumber\\
\ &=& -\sum_{J} \Lambda_{JI} n_J\ ,
\label{anarg}
\end{eqnarray}
where the $n_J$ are integers due to the topological quantization of the
electric field in $(1+1)$ dimensions, and represent the number of
{\it vortices} (quasi-particles) created in the $I$-th component of the fluid.
This spectrum of $\left( V^0_0 \right)^{I}$ must coincide with the
spectrum given by the lattice (\ref{lat}), which identifies
$\Lambda_{iJ} \equiv (v_i)^J$.
We have shown that, given the lattice $\Gamma$, the anomaly fixes the
definition of the physical charge (\ref{physch}), which in turn implies an
``interaction'' among the edges with respect to the diagonal basis
(\ref{cqj}).

The spectrum of physical charges follows directly from
(\ref{physch}) and~(\ref{anarg}),
\beq
Q\ \equiv\ \sum_{I=1}^m\ \rho_0^I\ =\ \sum_{i,j=1}^m
\ K^{-1}_{ij} \ n_j\ ,
\label{chspec}
\eeq
where we call $\ K^{-1}_{ij} = \sum _I v_i^I v_j^I \ $
the {\it metric} of the lattice.
The spin spectrum follows similarly from~(\ref{cqj}) and~(\ref{lat}):
\beq
J={\theta\over 2\pi}= {\frac{1}{2}} \sum_{i,j=1}^m\ n_i \ K^{-1}_{ij} \ n_j\ .
\label{jspec}
\eeq
Equations~(\ref{chspec}) and~(\ref{jspec}) give us the charge, spin and
statistics of the quasi-particle excitation corresponding to any point of
the lattice, parametrized by the vorticity components $n_i \in {\bf Z}$.

Finally, the filling fraction $\nu$ is easily computed by
applying a homogeneous tangential electric field
$E^I = E$, for all $I$.
The rate of total physical charge created by the anomaly on the edge is
$ \partial_t \left(-Q\right) =-E\ \sum_{ij} K^{-1}_{ij} \ ,$
which corresponds to a radial flux of charge, {\it i.e.}, to the Hall
current \cite{Stone}\cite{Cap2}. This identifies the filling  fraction as
\begin{equation}
\nu=\sum_{i,j=1}^m\ K^{-1}_{ij}\ .
\label{nu}
\end{equation}

We thus reach the conclusion that all two-dimensional incompressible
quantum fluids are classified by an integer level $m$ (the central
charge of $\ W_{1+\infty}\ $) and a real positive-definite symmetric
matrix $K^{-1}$ describing the metric of the representation lattice.
Note that our approach cannot predict gaps and excitation energies.
Nevertheless, a stability principle based on maximal symmetry of $\ K\ $
selects the most prominent fractions observed in experiments \cite{Cap3}

In applying the results (\ref{chspec})-(\ref{nu}) to the quantum Hall effect,
we should further require the presence of $m$ excitations with unit
charge, fermionic statistics and integer statistics relative to
any other excitation, representing the original electrons
in each of the $m$ components of the fluid.
This imposes that $K$ has {\it integer} entries, {\it odd} on the diagonal
\cite{Zee}.

The results (\ref{chspec})-(\ref{nu}) are in agreement with the
general hierarchy obtained by Fr\"ohlich, Wen and Zee~\cite{Zee}
by an Abelian Chern-Simons effective field theory.
Examples of $K$-matrices for the best known filling fractions have been
discussed by these authors; we refer to them
for a full discussion of the physical consequences of
eq.(\ref{chspec})-(\ref{nu}).

While following the same physical approach
of describing the universal long-distance properties,
the methods of ref. \cite{Zee} differ substantially from ours.
They started from an appropriate choice of effective field theory,
whereas we derived our results from purely algebraic considerations,
after having identified the {\it symmetry principle} governing
the phenomenology of the quantum Hall effect.
A direct relation between the two approaches can be found {\it a posteriori}.
The $(1+1)$-dimensional $m$-component chiral boson,
compactified on the torus ${\rm T}^m ={\bf R}^m/\Gamma$,
gives an explicit realization \cite{Cap3} of all (rational) $\winf$ theories,
and it is also
the edge degree of freedom of the Abelian Chern-Simons topological
theory on a disk \cite{Wen}.

Moreover, our classification shows that the results
(\ref{chspec})-(\ref{nu}) are {\it complete}.
We can in fact exclude further possible effective theories:
orbifold compactifications of the chiral boson \cite{Gin} lead to
excitations with ill-defined charge $V^0_0$;
non-Abelian Chern-Simons theories possess excitations with
non-Abelian statistics which are described by edge theories
with non-integer central charge \cite{Wen}.
Both possibilities cannot realize the $\winf$ symmetry.

We conclude by remarking that the highest weights for all $\winf$ currents
measure the higher moments of the charge distribution of excitations.
The fact that these are completely specified as polynomials of the charges
\cite{Kac}, has to be regarded as a further consequence of incompressibility.

We thank P.Sorba for bringing the paper by V.Kac and A.Radul
to our attention and V.Kac for useful explanations.
We also thank CERN for hospitality.

\end{document}